# Tenure Under Pressure:

# Simulation of AI's Disruptive Effects on Academic Publishing

Shan Jiang

University of Massachusetts Boston

**Abstract**

Generative artificial intelligence (AI) has begun to reshape academic publishing by enabling the rapid production of submission-ready manuscripts. While such tools promise to enhance productivity, they also raise concerns about overwhelming journal systems that have fixed acceptance capacities. This paper uses simulation modeling to investigate how AI-driven surges in submissions may affect desk rejection rates, review cycles, and faculty publication portfolios, with a focus on business school journals and tenure processes. Three scenarios are analyzed: a baseline model, an Early Adopter model where a subset of faculty boosts productivity, and an AI Abuse model where submissions rise exponentially. Results indicate that early adopters initially benefit, but overall acceptance rates fall sharply as load increases, with tenure-track faculty facing disproportionately negative outcomes. The study contributes by demonstrating the structural vulnerabilities of the current publication system and highlights the need for institutional reform in personnel evaluation and research dissemination practices.

## 1. Introduction

The rapid rise of generative artificial intelligence (AI) is reshaping academic work in profound ways. AI systems are now capable of producing full manuscripts, complete with citations, figures, and statistical analyses, at a fraction of the time and effort required by traditional scholarship (Yamada et al., 2025; Miao et al., 2025; Moss, 2025; Xu, 2025). Reports of "automated scholar AI" tools, such as Sakana AI, illustrate how research writing could be scaled almost exponentially, raising urgent questions about how existing publication and evaluation systems will respond (Lu et al., 2024; Castelvecchi, 2024). Early evidence from computer science conferences such as AAAI and NeurIPS, which have experienced unprecedented surges in submissions[1][2], suggests that these disruptions are no longer speculative but already unfolding (Kim et al., 2025).

For business schools and related disciplines, the implications are especially acute. Tenure and promotion decisions remain heavily reliant on publication counts in established journals (Dennis et al., 2006; Burton et al., 2023). Yet if AI accelerates manuscript production while journals maintain fixed acceptance capacities, faculty may face dramatically higher desk rejection rates, longer delays from submission to acceptance, and shrinking portfolios of published work. These structural pressures are likely to affect junior faculty most, as their careers depend on reaching publication benchmarks within six-year probationary periods. Tenure and promotion committees typically evaluate candidates against the publication patterns of recent cohorts, using the most current acceptance rates and journal submission volumes as implicit benchmarks (Tripodi et al. 2025). While this practice has been reasonable in the past, it could be problematic in the upcoming years because conditions in the publication market are changing dramatically due to the AI and its impact on publication market. If junior faculty were judged against performance standards set under more favorable publishing conditions, it would create criteria that unfairly penalize them despite equivalent or greater effort.

---

[1] https://aaai.org/conference/aaai/aaai-26/review-process-update/
[2] https://www.ctol.digital/news/ai-research-summit-neurips-2025-receives-record-breaking-27000-paper-submissions

This paper uses simulation modeling to explore these systemic vulnerabilities. By constructing a baseline model of faculty productivity and journal review processes, and extending it to scenarios involving early AI adopters and widespread "AI abuse," we assess how submission surges alter desk rejection behavior, review times, and eventual acceptance probabilities and faculty research portfolio at the end of tenure clock. Our analysis highlights the role of editorial triage as the most immediate bottleneck and illustrates how publication-based evaluation systems may become increasingly misaligned with the realities of AI-driven scholarship.

Ultimately, this paper contributes to emerging conversations on AI, academic publishing, and tenure evaluation. It provides a structured framework for anticipating risks, guiding institutional reform, and encouraging broader debates about how knowledge should be disseminated and recognized in the age of AI.

## 2. Background

### 2.1. Recent Surge of AI Papers

The recent acceleration of generative AI technologies has led to a striking surge in AI-assisted and AI-generated manuscripts across multiple domains. While human–AI co-authorship has existed since the early adoption of tools like GPT-3, the past two years have introduced far more sophisticated systems capable of automating entire stages of scholarly production. Most notably, *Sakana AI*[3] has emerged as an "automated scholar AI," demonstrating how large-scale language models can be combined with scientific workflows to generate complete manuscripts, including code, figures, data analyses, and references (Lu et al., 2024; Castelvecchi, 2024; Yamada et al., 2025). Other experimental systems have been reported that promise to "scale research" by producing hundreds of drafts, some of which appear submission-ready in hours rather than the months or years usually required (Jansen et al., 2025; Team et al. 2025). These developments have intensified academic and public debates on whether the publishing system can sustain its current evaluation standards when exposed to exponential increases in submissions[4] (Örtenbla and Koris, 2025).

---

[3] https://sakana.ai/ai-scientist/
[4] These developments have intensified academic and public debates on whether the publishing system can sustain its current evaluation standards when exposed to exponential increases in submissions.

Evidence of this disruption is already visible in fields that are closest to AI development. Computer science conferences such as AAAI, NeurIPS, and ICML have reported dramatic increases in submission counts (see footnote in page 1), with AI writing tools widely suspected of further inflating volumes (Kim et al., 2025). Organizers and program committees now openly discuss the unsustainable burden this places on reviewers[5], who are often overwhelmed by both the sheer quantity of papers and the difficulty of distinguishing between genuinely novel contributions and AI-assembled content (Chen et al., 2025).

While computer science has served as the first testing ground, scholars in other disciplines are increasingly aware that their fields will not remain insulated. Editors across domains, including business, economics[6], and nature science (Adan 2025), have begun voicing concerns about "submission flooding," even if most evidence so far remains anecdotal. What is clear, however, is that the rate of manuscript generation has outpaced the linear capacity of editorial and peer review processes. Some have warned that without structural changes, academic publishing could face a "capacity crisis," where acceptance rates plummet not due to higher quality standards but because of overwhelming demand (Horta and Jung, 2024). As a result, journals, institutions, and individual scholars have begun debating whether new evaluation criteria, technological interventions, or alternative dissemination platforms are needed to maintain fairness in scholarly recognition (Aczel et al., 2025; Bai et al., 2025).

## 2.2. Triage Processes and Desk Rejection

The triage or desk rejection stage plays a central role in scholarly publishing, particularly in high-prestige journals where acceptance rates are low. At this stage, the Editor-in-Chief (EIC) or senior editorial board members make rapid decisions about whether a submission should proceed to full peer review. Desk rejection typically accounts for 40–70% of decisions in many business and management journals, and rates are often even higher in top-tier outlets (Bhukya et al., 2022; Beugelsdijk and Bird, 2025). The decision process is necessarily selective and frequently based on limited information, primarily the title, abstract, cover letter[7], and sometimes a quick scan of the introduction and methodology (Beugelsdijk and Bird, 2025). Editors seek to judge both fit (scope and relevance to the journal's audience) and baseline quality (clarity, novelty, methodological soundness). Papers deemed weak or misaligned are quickly removed from the pipeline, conserving reviewer resources for more promising manuscripts.

This system, however, is particularly vulnerable to AI-driven shocks. First, desk rejection is already a manual and labor-intensive process[8], demanding that editors personally screen large volumes of manuscripts under tight time constraints. As submissions double or triple in volume due to AI-boosted productivity, editorial workloads become unsustainable, raising the risk of errors. Second, because triage decisions are made with partial information, higher volumes magnify the likelihood of false negatives (Dadhich et al., 2025), cases where innovative work is rejected simply because editors cannot invest enough time in deeper evaluation. This problem disproportionately affects early-career researchers or those working at less prestigious institutions, as editors often unconsciously rely on heuristics such as author reputation, institutional affiliation, or familiar methodologies when overwhelmed (Lee et al., 2013; Tomkins et al., 2017).

---

[5] https://blog.neurips.cc/2025/05/02/responsible-reviewing-initiative-for-neurips-2025/
[6] https://blogs.worldbank.org/en/impactevaluations/long-slog-publishing-economics
[7] https://ecrlife.org/why-desk-rejections-happen/?utm_source=chatgpt.com
[8] https://www.linkedin.com/posts/jason-thatcher-0329764_academicpublishing-deskrejects-editorperspective-activity-7324252555759259649-4qaF

Moreover, the desk rejection stage serves as a gatekeeper for reviewer capacity. Reviewer availability has already become a crisis: surveys indicate that more than half of review invitations are now declined[9], and editors frequently report difficulty in securing even two reviewers (Horta and Jung 2024; Thompson et al., 2024). Many reviewers start to express fatigue for unpaid peer review tasks (Beecher and Wang, 2025). Since reviewers cannot scale linearly with submissions, surges in manuscript volume lead editors to respond by increasing desk rejection rates to maintain manageable review pipelines (Beugelsdijk and Bird, 2025).

While AI-assisted triage tools have been proposed, such as automated screening for scope fit, plagiarism, or methodological red flags (see e.g., Checco et al., 2021; Santra and Majhi, 2023), many journals remain reluctant to adopt them, citing concerns about bias, transparency, and the ethics of delegating editorial judgment to machines[10][11]. As a result, the mismatch between exponentially growing submissions and linear human editorial capacity is likely to become more notable. Desk rejection may thus become the most visible bottleneck as AI accelerates scholarly output.

## 3. Methods

### 3.1 Rationale for Simulation

This study employs simulation as the primary methodological approach to investigate the impact of AI-driven submission surges on journal editorial processes and their downstream consequences for faculty tenure processes. Simulation is particularly appropriate in this context for three reasons.

First, many of the relevant metrics in the publication pipeline are not publicly transparent (Hanson et al., 2024; Sayab et al., 2025). Data such as journal desk-rejection probabilities, editorial triage times, reviewer acceptance rates, and personnel-level tenure outcomes are typically internal to journals or institutions. Simulation, while not yielding precise estimates, provides a backbone understanding of system dynamics. The outputs can serve as baseline scenarios that can later be adjusted or calibrated by journal offices, academic departments, or tenure committees for their specific circumstances.

Second, the tenure review process in many business schools relies on comparison with prior cohorts rather than contemporaneous peers, since often only one junior faculty member is reviewed within a department at a given time. This practice creates a structural risk of unfairness: if systemic publication delays caused by AI-driven submission surges are not quantified, junior faculty may be disadvantaged relative to earlier cohorts, even when their effort and quality are equivalent.

Third, although survey evidence can provide insights into evolving practices, large-scale surveys are prone to bias. Junior faculty may have incentives to overstate difficulties to justify publication shortfalls, while senior faculty may understate challenges to protect the legitimacy of existing systems (Jong and Kantimm, 2024; Launio et al., 2024). For this reason, we use small-scale survey findings and anecdotal evidence primarily to inform model design, while relying on simulation to explore systemic consequences.

---

[9] https://www.insidehighered.com/news/2022/06/13/peer-review-crisis-creates-problems-journals-and-scholars
[10] https://scholarlykitchen.sspnet.org/2025/09/17/peer-review-in-the-era-of-ai-risks-rewards-and-responsibilities/
[11] https://asm.org/articles/2024/november/ai-peer-review-recipe-disaster-success

### 3.2 Behavioral Assumptions in Baseline Model

**Faculty Strategy**

In the baseline simulation, each faculty member produces a steady flow of manuscripts at a fixed productivity rate (e.g., one or two papers per year). The number of manuscripts generated each year follows a Poisson distribution, ensuring that research output varies realistically across individuals and years. Submissions are spread across a six-year horizon to mirror common academic practice, where projects are initiated and submitted continuously.

These manuscripts are submitted following a tiered submission ladder (T1->T2->T3), where authors typically begin with higher-tier outlets and, if unsuccessful, move stepwise toward lower-tier journals (Calcagno et al., 2012; Kovanis et al., 2017). When a manuscript is desk rejected, the faculty member immediately resubmits the same manuscript to another journal within the same tier, reflecting the view that desk rejection often indicates non-fit with a specific outlet rather than overall quality (Beugelsdijk and Bird, 2025). By contrast, if a manuscript is rejected after peer reviews, the author interprets this outcome as an evidence that the paper may not meet standards of the journal in the current tier. In this case, the manuscript is resubmitted to another journal one tier lower in the ladder. However, in case a manuscript keeps getting desk-rejected by journals in the same tier, the authors would usually go down one tier in reality because it also signals weak quality of paper (Rutherford, 2025). Therefore, we assume that after three desk rejections, a manuscript is resubmitted to the next tier. For Tier-3 journals, a manuscript is considered to fail after three desk rejections.

If a manuscript receives a major revision, the author undertakes the revision process and resubmits to the same journal for further reviews. This cycle continues until the manuscript is either accepted, rejected by Tier 3 journals for three times, or the six-year horizon ends. The baseline scenario assumes no acceleration from AI tools, focusing instead on pre-AI era faculty productivities and standard journal workflows. However, we will model the impact of AI as an external influx of paper submissions to the simulation system in extended models that will be discussed later.

**Editor Strategy**

In our simulation, we explicitly model the editor's decision-making process, as desk rejection is directly controlled by editors (especially editors-in-chief and senior editors) and is expected to be the most immediate bottleneck under AI-accelerated submission growth. Editors act as gatekeepers: upon receiving a manuscript, they either desk reject it or send it to review, following tier-specific probabilities.

Once a manuscript enters review, the process is modeled in discrete rounds. In each round, the editorial decision can be:

- Accept (with increasing probability in later rounds),

- Major Revision (with declining probability across rounds), or
- Reject (ending the process at that journal).

This structure reflects how editors and associate editors typically combine referee feedback with their own judgment: acceptance is uncommon in early rounds, while later rounds tend to converge toward a final decision (Huisman and Smits, 2017). We set a maximum of three review rounds (K=3). If a manuscript is still under major revision after the final round, it is treated as accepted. This assumption serves two purposes: it keeps the model parsimonious, and it reflects a "kind" academic world in which repeated revisions improve manuscript quality and ultimately lead to publication rather than indefinite cycles (Nederhof, 2006; Powell, 2016).

We do not model reviewers as individual agents. Instead, reviewer contributions are encapsulated in review cycle-level parameters including review round length, round-specific acceptance rates, and major revision probabilities. This design maintains model parsimony while still capturing the aggregate dynamics of the peer-review process. Although AI may eventually influence reviewers (e.g., faster report preparation, automated checks), its systemic impact is expected to fall most heavily on editors, who must triage and process dramatically larger submission volumes. Our baseline model therefore focus on editorial-level strategies.

Combined together, the following pseudo code represents the interaction between a faculty and editors of multiple journals.

```
For EACH year, draw a manuscript count from a Posson distribution
  For EACH manuscript, set submission start month drawn from a uniform distribution from the year.
    START at T1 journal, time = 0
    WHILE paper not accepted and paper stay within T1, T2, T3:
      ADD desk decision time;
      IF desk rejected (<3 times) → stay in same tier, CONTINUE
        IF desk rejected the 3rd time, move to next tier, CONTINUE
      FOR each review round:
        ADD journal review time (drawn from truncated normal distribution)
        IF accepted (or R3 major revision) → STOP
        IF major revision in R1/R2 → ADD author revision time (drawn from truncated normal distribution), CONTINUE round
        IF rejected → move to next tier, BREAK
    END
```

### 3.3 Baseline Model Parameters

Table 1 lists the baseline model parameters. We explain the rationale for setting the initial parameter values next.

Table 1. Summary of Baseline Model Parameters

| Parameter | Initial Value(s) | Rationale | Source / Citation |
|---|---|---|---|
| [1]Environment | Faculty Pool size=30,000<br>Journal=100<br>Time Horizon=6 years | An overall estimate of real-world simulation of business school publication market | AACSB (2025)[12] |
| [2]Journal tiers | T1=7%<br>T2=24%<br>T3=69% | Approximation of A*/A/B–C distribution in business school journals | ABDC Journal Quality List (2022)[13] |
| [3]Desk-reject probability | T1=0.70<br>T2=0.50<br>T3=0.30 | Anecdotal reports that top journals desk reject >50% of submissions; developmental outlets lower | Elsevier Connect (2015)[14]<br>Charlesworth Author Services (2022)[15] |
| [4]Desk-decision time | 0.3 mo. | Median of authors' reported desk rejection time | SciRev (2025)[16] |
| [5]Review times per round | T1=6.0 mo. T2=5.0 mo. T3=4.0 mo. With SD | Typical editorial cycle lengths. Numbers are mean of normal distribution. See Appendix A for SDs | Publons (2018), SciRev (2025) |
| [6]Author revision times per round | T1=2.5 mo. T2=2.0 mo. T3=1.5 mo. | Time needed by authors to address major revision comments. Numbers are mean of normal distribution. See Appendix A for SDs | Heuristics |
| [7]Acceptance rate | T1=0.09<br>T2=0.12<br>T3=0.24 | Eventual acceptance rate of journals | ABDC Journal Quality List (2022), Selected Journal Websites. See Appendix B. |

---

[12] https://www.aacsb.edu/insights/reports/2025/2025-business-school-data-guide
[13] https://abdc.edu.au/abdc-journal-quality-list
[14] https://www.elsevier.com/connect/5-ways-you-can-ensure-your-manuscript-avoids-the-desk-reject-pile
[15] https://www.cwauthors.com/article/reasons-for-desk-rejection-and-how-to-avoid-desk-rejection
[16] https://scirev.org/

| Parameter | Initial Value(s) | Rationale | Source / Citation |
|---|---|---|---|
| [8]Per-Cycle Acceptance and Revision probability | Varies by Journal Tier and Review Rounds | Encodes multiple revision rounds as common | Reversely estimated from [3] and [7]. See Appendix C. |
| [9]Resubmission gaps | After desk reject=0 mo. After review reject=0 mo. | Optimistic assumption. Faculties immediately resubmit to race for tenure clocks. | Heuristics |

**Environment**

In this simulation, we fix the environment at 100 journals to capture the scale of the common journal lists used by business schools, while keeping the model tractable. Specifically, the Australian Business Deans Council journal list (ABDC 2022) is used as the source for estimating model parameters. The ABDC Journal Quality List is more suitable for this study than the FT50 or ABS lists because of its breadth, global reach, and tiered structure. ABDC includes more than 2,800 journals spanning all business disciplines and categorizes them into A*, A, B, and C tiers, which is essential for modeling variations in acceptance rates, desk rejections, and review times. The number and journal variations are reduced to 100 abstract journals (T1, T2, T3 tiers) for computational efficiency.

On the faculty side, we set the pool at 30,000 research-active business faculty worldwide. To assess the plausibility of our assumptions, we conducted a simple face-validity check of submission volume relative to journal capacity. With approximately 30,000 research-active faculty and an average productivity of two papers per year, the system produces roughly 60,000 submissions annually. Distributed across the 100 journals modeled in our simulation, this equates to about 600 submissions per journal per year. In practice, most business and management journals publish only 50–120 articles annually, depending on issue size and frequency. This implies overall acceptance rates in our model fall in the range of 8–20 percent, which is in the realistic range.

For duration, we adopt a six-year simulation horizon because it mirrors the typical probationary period for junior faculty in business schools, during which publication output is a primary criterion for tenure and promotion decisions.

**Journal Tiers**

In our simulation, journals are categorized into three tiers that mirror the ABDC Journal Quality List (2022) distribution. We assign 7% of outlets to Tier 1 (A*), representing flagship journals with the lowest acceptance rates and longest review cycles; 24% to Tier 2 (A), covering strong disciplinary journals with moderate selectivity; and the remaining 69% to Tier 3 (B/C), which reflects the bulk of outlets offering broader or developmental publication opportunities. This tiered structure is critical because journal behavior varies systematically across categories: A* journals are known for high desk rejection probabilities and extended timelines, A journals offer more balanced though still competitive review processes, and B/C journals tend to have higher acceptance probabilities and shorter cycles. By aligning our model with the actual ABDC tier proportions, we ensure that simulated outcomes realistically capture the uneven distribution of prestige, selectivity, and editorial capacity across the publication market.

**Desk Reject Probability and Time**

In setting the parameters for desk rejection probability and time, our baseline model relies on a combination of official sources and complementary evidence from SciRev downloaded data. We begin with the recognition that journals, particularly at the A* and A levels, routinely report high desk-reject rates due to volume pressures and fit considerations. Editor statements and journal blogs often indicate desk rejection rates of 50–70% for top-tier outlets and somewhat lower but still substantial rates for mid-tier journals (see e.g., Elsevier Connect, 2015; Charlersworth Author Services, 2022). Based on this evidence, we assign baseline probabilities of 0.70 for Tier 1, 0.50 for Tier 2, and 0.30 for Tier 3, reflecting the stricter triage practices at more prestigious journals. For desk rejection time, we first incorporate insights from the "immediately rejection" field in the SciRev database, where the median reported delay is approximately 0.3 months (9 days). This aligns with editorial claims that desk decisions are intended to be fast, even if authors sometimes experience longer waiting times.

Importantly, in the baseline model we do not yet incorporate AI's potential to accelerate or complicate desk rejection processes. Instead, we calibrate against current practice, using official editor reports where available and drawing on SciRev author-reported timelines as a supplemental anchor when official data are missing.

**Review Time**

For the review cycle time parameter, we distinguish journal tiers because review duration is closely tied to journal selectivity. In our baseline model, we set review cycle times at 6 months for Tier 1 (A*), 5 months for Tier 2 (A), and 4 months for Tier 3 (B/C). These values reflect both public report (Publon 2018) and empirical evidence. Major publishers such as Elsevier and Springer Nature note that peer review typically ranges from 3 to 6 months, depending on field and journal selectivity. Highly prestigious journals, such as those in the A* category, often report longer turnaround times due to difficulty in securing reviewers and the detailed nature of referee reports, making 6 months a conservative yet realistic choice. For Tier 2 and Tier 3 journals, slightly shorter review times are common, hence the 4-5 month setting. These values are further supported by aggregate data from SciRev, where total handling times often span 12–18 months across multiple rounds, implying that a single round typically lasts 4–6 months.

**Acceptance Rate**

Acceptance rate parameters in the simulation are calibrated using a combination of official journal-reported data and field-level adjustments. For Tier 1 (A*), we set acceptance at 9%, reflecting the consistently low rates reported by flagship outlets. For Tier 2 (A), we assign 12%, based on reported ranges across strong disciplinary journals that are somewhat more accessible but still selective. For Tier 3 (B/C), we set 24%, acknowledging the broader coverage of journals, their more developmental orientation, and the tendency for authors to target these outlets after higher-tier rejections. These estimates are grounded primarily in information available on official journal or publisher websites. While the data are not available at every journal, we selected representative journals where such data is available to calculate the estimates. A full documentation of sources and supporting details is provided in Appendix B.

**Revision Probability**

Each review round can lead to one of three outcomes: acceptance, revision, or rejection. This structure reflects real editorial practice, where manuscripts may occasionally be accepted early, more commonly undergo one or more rounds of major revision, or be rejected outright.

Including explicit revision probabilities ensures that acceptance is not modeled as a single-step event but as the outcome of a sequence of editorial decisions across multiple rounds. This design is essential for capturing the additional time incurred through repeated cycles of revision, resubmission, and re-review, which substantially influences the total waiting time to final publication.

Because direct data on per-round outcomes are rarely available, we calibrate these probabilities indirectly. Starting from each tier's overall acceptance rate and desk rejection probability, we heuristically estimate round-specific acceptance, revision, and reject probabilities that reproduce observed eventual acceptance rates. The details of this calibration are documented in Appendix C.

**Author Revision Time and Resubmission Gaps**

Finally, we include two heuristic-based parameters in the model. Author revision time is incorporated into the model to account for the delay introduced when authors respond to major revision requests. Unlike journal review cycles, which are partially documented in publisher reports or databases, author revision times are rarely tracked systematically and vary widely depending on the scope of requested changes, teaching schedules, and other personal constraints. For this reason, we rely on heuristic estimates informed by publisher guidance, which commonly specifies two to three months for major revisions. In our model, we set 2.5 months for Tier 1, 2.0 months for Tier 2, and 1.5 months for Tier 3, reflecting the expectation that more prestigious journals demand more extensive revisions. While heuristic, these values are overall consistent with editorial practice and ensure that our timelines remain realistic.

Resubmission gaps capture the downtime between rejection and resubmission to another outlet. Since these delays depend on individual author behavior and strategy rather than formal journal processes, they are rarely observable in any dataset. To keep the model tractable, we set the resubmission gap to 0 months for both desk rejections and rejections after review, representing the best-case scenario in which authors immediately reformat and resubmit without delay. This assumption underestimates real-world lags but ensures the simulation reflects the most efficient possible strategies for authors. Consequently, our results may slightly underestimate the impact of AI on publication timelines, as actual resubmission processes often take longer.

## 4. Results

### 4.1 Baseline Model Evaluation

In this study, we carefully designed a baseline simulation model and parameterized it to approximate the dynamics of the real-world publication market. To further assess the validity of the baseline model, we conducted two complementary analyses using the initial parameter settings, focusing on key outcome metrics including (1) total time to publication, and (2) total number of acceptance during faculties' six-year tenure clock.

**Submission to Acceptance Analysis**

In submission-to-acceptance analysis, we focus on how long a single manuscript takes from its initial submission until final acceptance. 10,000 submission were simulated and submitted to journals in our model. The results are presented in Figure 1.

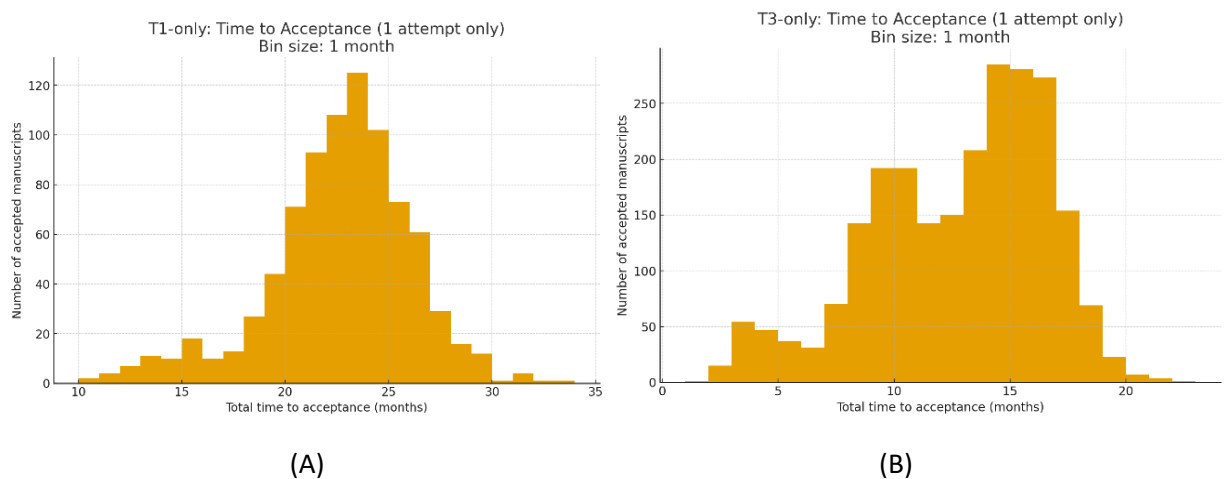

| (A) | (B) |
|---|---|
| **Accepted:** 843 | **Accepted:** 2,380 |
| **Desk-rejected:** 7,007 | **Desk-rejected**: 2,978 |
| **Review-rejected**: 2,150 | **Review-rejected:** 4,642 |
| **Overall acceptance rate**: 8.43% | **Overall acceptance rate**: 23.8% |
| **Median time to acceptance**: $\approx$ 23.04 months | **Median time to acceptance**: $\approx$ 13.65 months |

Figure 1 Time to Acceptance Analysis

In analysis (A), all manuscripts were submitted to Tier 1 (T1) journals. Rejected papers, whether by desk screening or peer review, were not resubmitted. Out of 10,000 submissions, 843 were ultimately accepted, consistent with the pre-set T1 acceptance rate. The median time to acceptance was 23.04 months, and the distribution of total time to acceptance is shown in Figure 1(A).

In analysis (B), all manuscripts were submitted to Tier 3 (T3) journals, again without resubmission after rejection. Among 10,000 submissions, 2,380 were accepted, matching the pre-set T3 acceptance rate. The median time to acceptance was 13.65 months, as shown in Figure 1(B). Overall, the time to acceptance in T3 journals was roughly 60% of that in T1 journals.

We further examined a "submit-until-acceptance" (SUA) model (Figure 2). In this model, manuscripts are first submitted to T1 journals. Papers rejected after peer reviews are resubmitted to the next tier (e.g., from T1 to T2), while desk-rejected papers are resubmitted within the same tier for at most three times. Under these assumptions, 2,855 papers were ultimately accepted at T1, 1,704 at T2, and 5,441 at T3. The median time to acceptance was 30.54 months.

Overall, these simulation results aligned with our intuition. In the SUA model, the number of papers accepted at T2 journals was lower than expected, reflecting the assumption that all manuscripts start at T1. In practice, however, many authors target T2 journals as their first choice.

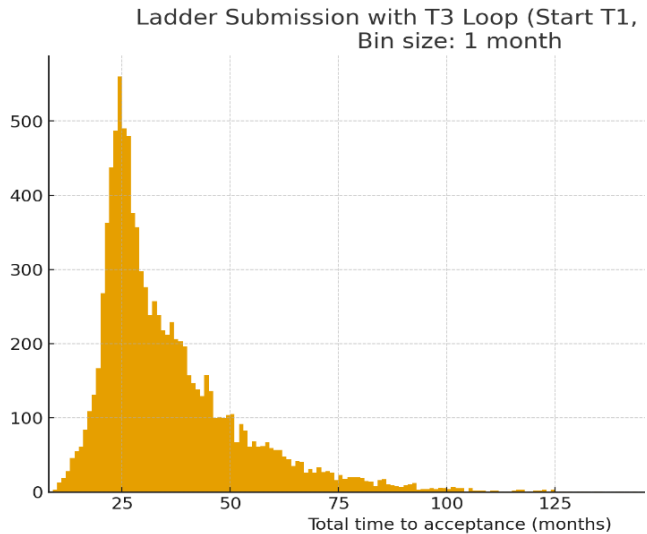

Figure 2 Total Time Taken If Submit Until Acceptance

**Faculty Tenure Portfolio Analysis**

In Faculty Tenure Portfolio analysis, we focus on how many papers an author can get accepted during the six-year time frame (tenure clock). We still use assume that facultries will start from T1 journals, and resubmit to next journal tier in the ladder whenever review-rejected. For faculty productivity, we varied the the faculty productivity and changed number of submitted manuscripts per year from 1 to 2, representing typical productivity of research-active faculties. We ran simulation models with 30,000 faculties. Figure 3 shows the results.

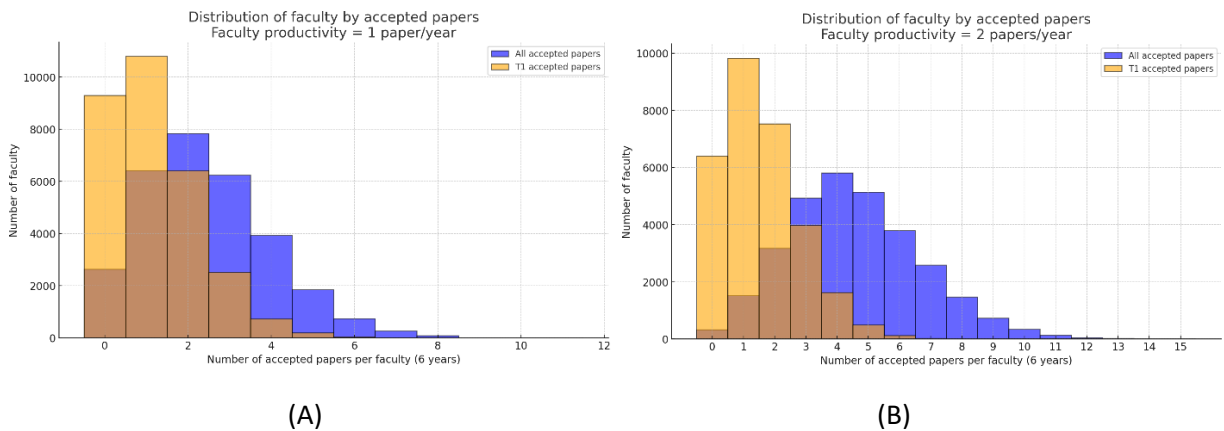

(A)                       (B)

| | |
|---|---|
| **Faculty Pool Size**: 30,000<br>**Faculty Productivity**: 1 paper per year<br>**All accepted papers**<br>Mean = 2.43; Median = 2; Std. dev. = 1.55<br>**T1 accepted papers**<br>Mean = 1.18; Median = 1; Std. dev. = 1.09<br>**Total accepted**: 72,851<br>  **Total accepted at T1**: 35,327<br>**Total desk rejections encountered**: 554,054 | **Faculty Pool Size:** 30,000<br>**Faculty Productivity:** 2 papers per year<br>**All accepted papers**<br>Mean = 4.55; Median = 4; Std. dev. = 2.13<br>**T1 accepted papers**<br>Mean = 2.56; Median = 1; Std. dev. = 1.25<br>**Total accepted:** 136,350<br>  **Total accepted at T1:** 46,706<br>**Total desk rejections encountered:** 853,630 |

Figure 3 Faculty Research Portfolio Analysis

When assuming a productivity rate of one paper per year, most faculties published 1 or 2 T1 journal papers within six years. About 17% did not publish in any Tier 1 journal within six years, while approximately 26% published more than two papers in Tier 1 journals. The total number of accepted papers was mostly between 2 and 5.

When assuming a productivity rate of two papers per year, about 20% of the faculties did not publish in a Tier 1 journal within six years, while roughly 33% published one Tier 1 paper, and 25% published two papers in Tier 1. The total number of accepted papers were mostly between 2 and 7.

Overall, the simulation results are consistent with general expectations about the global publication market. We acknowledge, however, that publication profiles can vary substantially across individual faculty, institutions, and national contexts. The aim of this model evaluation is not to claim that the framework precisely reflects real-world dynamics, but rather to suggest that institutions and policy makers whose faculty performance patterns resemble those captured in the simulations may find the insights from this study particularly relevant and useful.

**4.2 Impact of AI**

From this subsection, we will extend the baseline simulation model to examien the impact of AI on journal publication system, and further its impact on tenure system. We assume faculty productivity = 2 papers per year in this section of analysis because it better represents the competitive publication market in US.

**Early Adopter Model**

In this scenario, we assume that 10% of faculty adopt AI tools and increase their productivity to 20 papers per year, while the remaining 90% maintain the baseline rate of 2 papers per year. This results in roughly 60,000 additional manuscript submissions per year entering the publication market. While such a surge in submissions will affect multiple stages of the publication process, our analysis focuses specifically on desk rejection, as it serves as the primary gateway for handling the influx.

The impact of increased submissions on journal offices can be understood in two ways:

1. **Longer triage times**, since editors must screen more papers.

2. **Higher desk rejection rates**, in order to keep the number of papers sent out for review roughly constant, given the growing difficulty of securing reviewers.

For this analysis, we focus on the change in desk rejection rates while holding triage time constant. There are two reasons for this choice. First, the baseline desk rejection time in our model is already very small (0.3 months). Even if this value increased linearly, the effect on a six-year horizon would be marginal compared to the effect of higher rejection rates. Second, in the long run, it is unlikely that editors will continue to extend triage times indefinitely. As AI-driven submissions grow, editorial offices are expected to adopt AI-assisted triage, which would stabilize or even reduce desk rejection time (Wei et al., 2025).

To capture the effect from increased desk rejection rate, we introduce the concept of external load (L). The desk rejection rate p(L) at load level L is defined as:

$$p(L) = 1 - \frac{1 - p(\text{baseline})}{L}$$

This formula operationalizes the editorial strategy of keeping the number of manuscripts sent for review approximately stable, even under dramatically higher submission volumes. For example, if desk acceptance was 30% at baseline and submissions double (L=2), then only 15% of papers will now pass desk review, effectively holding constant the absolute number of papers entering the review process.

In the Early Adopter Model, L=2, which yields adjusted desk rejection rates of 0.85 (T1), 0.75 (T2), and 0.65 (T3). Using these parameters, we simulate the system over a six-year horizon and compare the outcomes for the two groups of faculty: early adopters versus the rest. Table 2 reports the results.

The results in Table 2 show that faculties maintaining two papers per year face a clear decline in research outcomes once AI-driven submissions double the system load. Their mean total accepted papers drop from 4.55 to 3.53 over six years, and the median falls from 5 to 3. However, the effect on T1 papers is relatively smaller: the mean decreases from 1.56 to 0.91, but the median remains at 1. This relative stability arises because T1 journals already had high baseline desk rejection rates, so the incremental tightening does not radically alter outcomes at this tier.

By contrast, the larger decline in overall acceptances reflects the sharper rise in desk rejection at T2 and T3 outlets, which previously absorbed many manuscripts rejected from T1 journals. In effect, lower-tier journals are now less able to function as safety nets, widening the gap between adopters and non-adopters. Meanwhile, early adopters dominate across all tiers, with a striking increase in both total and T1 acceptances, widening disparities within the faculty population.

It is important to note, however, that the early adopters' outcomes in this simulation are somewhat inflated compared to present reality. In practice, not all AI-boosted papers are entering the publication market, as many scholars hesitate to submit them due to ethical uncertainties. Still, the potential surge in submissions is not underestimated—AI's productivity gains are already evident across many sectors. Thus, this table should be read as a precautionary scenario, illustrating the possible trajectory if large-scale AI adoption materializes.

Table 2. Early adopter model results: the shock of AI on faculties keeping the same research practice

| Metric | Baseline | Early Adopters (10%) | Rest of Faculty (90%) |
|---|---|---|---|
| **Sample Size** | 30,000 | 3,000 | 27,000 |
| **Productivity (papers per year)** | 2 | 20 | 2 |
| **Mean accepted papers per faculty** | 4.55 | 35.33 | 3.53 |
| **Median accepted papers per faculty** | 5 | 35 | 3 |
| **Std. dev. of accepted papers** | 2.13 | 5.97 | 1.88 |
| **Mean T1 accepted papers per faculty** | 1.56 | 9.08 | 0.91 |
| **Median T1 accepted papers per faculty** | 1 | 9 | 1 |
| **Std. dev. of T1 accepted papers** | 1.25 | 3.10 | 0.95 |
| **Total accepted (all tiers)** | 136,350 | 106,004 | 95,269 |
| **Total T1 accepted** | 46,706 | 27,227 | 24,514 |
| **Total desk rejections** | 853,630 | 1,627,235 | 1,460,926 |

**AI Abuse Model**

In this model, we consider an even worse scenario when AI is further abused in academia — how the unchecked use of AI to mass-produce manuscripts could reshape the publication market and, in turn, tenure outcomes for faculty. Unlike the Early Adopter model, where only a subset of scholars dramatically increase productivity, here we treat the influx of AI-generated papers as an external system-wide load. This framing reflects the broader risk that once AI tools become normalized in research workflows, the overall submission volume will expand well beyond current baselines.

In the AI Abuse models, forecasting with external load factor $L > 2$ is reasonable. Recent reports of fully automated "scholar AIs" capable of producing draft manuscripts (Castelvecchi, 2024), as well as anecdotal accounts of researchers already experimenting with such tools, highlight the plausibility of exponential increases in submissions (Kim et al., 2025; Wei et al., 2025). The surge of AI-assisted submissions at major conferences, for example, the recent uptick noted at AAAI, offers an early signal of what journals may soon face. Combined with the influence of early adopters and the inevitability of journal offices themselves turning to AI for triage, it is likely that more scholars will adopt AI tools, whether voluntarily or out of competitive pressure.

For this reason, the AI Abuse model no longer divides the faculty pool into adopters versus non-adopters. Instead, we model AI-driven submissions as an external input to the publication market. Conceptually, this approach is equivalent to examining the 90% of the faculties in the Early Adopter model if our goal is to understand AI's aggregate impact on the faculty population who do not use AI.

We use the following external load factors, and corresponding desk-rejection rates by load (L):

- **L=2:** T1=0.85, T2=0.75, T3=0.65
- **L=3:** T1=0.90, T2=0.833, T3=0.767
- **L=5:** T1=0.94, T2=0.90, T3=0.86
- **L=10:** T1=0.97, T2=0.95, T3=0.93

These values illustrate how rapidly the desk rejection barrier rises as submission volume expands, leaving fewer manuscripts to proceed to review and acceptance. Table 3 presents results about the impact of AI on individual faculties

Table 3: AI Abuse and its impact on individual facultie acceptance in six years

| External Load | Mean (All) | Median (All) | SD (All) | Mean (T1) | Median (T1) | SD (T1) |
| --- | --- | --- | --- | --- | --- | --- |
| **1 (Baseline)** | 4.55 | 4 | 2.13 | 1.56 | 1 | 1.25 |
| **2** | 3.54 | 3 | 1.89 | 0.92 | 1 | 0.95 |
| **3** | 2.79 | 3 | 1.67 | 0.65 | 0 | 0.80 |
| **5** | 1.95 | 2 | 1.40 | 0.40 | 0 | 0.64 |
| **10** | 1.08 | 1 | 1.04 | 0.21 | 0 | 0.46 |

Table 3 illustrates the sharp erosion of faculty acceptance outcomes under increasing levels of AI-driven submission load. In the baseline environment (L=1), faculty members average 4.55 accepted papers over six years, including 1.56 T1 acceptances. However, as the external load doubles (L=2), these averages drop to 3.54 and 0.92, respectively. This downward trend accelerates as load intensifies: by L=5, the mean number of accepted papers falls to 1.95, with fewer than half a T1 acceptance per faculty on average. Most faculties won't be able to make any papers accepted at a T1 journal. At L=10, the publication system collapses entirely: most faculty secure just a single acceptance over six years, with T1 publications effectively vanishing (mean 0.21).

Both the mean and median results underscore how pervasive this effect would be: what begins as a manageable decline at L=2 quickly escalates into systemic suppression of faculty output at higher loads. The reduction in T1 outcomes is particularly alarming, as these journals are central to tenure evaluations in business schools. In sum, the table demonstrates that without systemic adjustments, AI abuse could dramatically endanger the ability of individual faculty to maintain even minimal research portfolios, fundamentally destabilizing current tenure expectations.

## 5. Discussions

**5.1 Implicit Assumptions in the Model**

The results generated by our simulation models appear strikingly alarming, and they diverge from the realities currently experienced in academic publishing. Rather than being read as literal forecasts of what will occur, their primary value lies in highlighting potential systemic vulnerabilities under a set of simplifying assumptions. These results must therefore be interpreted cautiously, with careful attention to the fact that the models implicitly assumed:

1. AI boosts productivity in generating *submission-ready* (not publication-ready) papers in the scale of 10 times and higher.

2. Faculties face zero delay when resubmitting rejected manuscripts, either desk-rejected or review-rejected.

3. Faculties resubmit to a lower tier after review rejection, whereas in reality they often spend extra time cycling within the same tier.

4. Journals maintain their current acceptance capacity without increasing publication slots.

5. All submissions are to be absorbed by a fixed pool of journals in the market (e.g., 100 journals in our model), but not anywhere else like preprint servers.

6. Tenure evaluation systems continue to emphasize publication counts as the primary metric.

7. Most faculty (except early adopters in our model) maintain their traditional productivity levels despite the availability of AI.

We discuss how each of these assumptions affect our interpretations of results next.

**5.2 AI Productivity Boost**

First, the most radical implicit assumption we made was that AI boosts productivity in generating submission-ready (not publication-ready) papers by at least 10 times. Whether this assumption underestimates or overestimates AI's effect on productivity is difficult to pin down. On one hand, the assumption may be conservative: there is mounting evidence that AI can already generate full manuscripts, including figures, citations, and statistical outputs, within hours (Yamada et al., 2025; Beel et al., 2025). Media reports and preprints describe agentic 'scholar AI' systems that can autonomously generate complete manuscripts and run large-scale experiment batches rapidly; in documented tests, teams submitted multiple AI-generated papers to a workshop, with at least one accepted[17]. On the other hand, the assumption could also be seen as an overestimate if we consider that "submission-ready" does not guarantee "publication-level" quality. Many AI-produced drafts still require substantial human oversight for novelty, coherence, and methodological rigor (Yamada et al., 2025; Jiang, 2025a), and many faculties remain hesitant to submit such work due to ethical or reputational concerns (Daoudi, 2025; Wells et al., 2025).

Thus, the figure is best understood as an precautionary scenario: it captures the potential scale of disruption if barriers to AI adoption fall and norms about its use in scholarly publishing loosen. The reality may be lower in the short term, but current evidence suggests that the long-run risks are at least as severe as our modeling assumes.

### 5.3 Faculties are not Robots

In our simulation models, we made two simplified assumptions about resubmission behavior that likely lead us to underestimate the time required for faculty to move from first submission to final acceptance. First, faculties were assumed to face zero delay when resubmitting rejected manuscripts. In reality, even desk rejections involve unavoidable delays (Huisman and Smits, 2017). Editors may take days or weeks to issue a decision, and authors must reformat and adapt their manuscripts for the next outlet. In reality, many desk rejection takes longer than the 0.3 month (= 9 days) that was assumed in our model based on SciRev data.

Second, faculties were assumed to resubmit to a lower tier after any review rejection, rather than spending additional time cycling within the same tier. In real world, for review-based rejections, additional time is often consumed as authors revise content, seek feedback, and select new journals. Moreover, our model assumes an immediate tier descent after review rejection, but in practice, many scholars first resubmit to journals of comparable tier. This strategy can introduce 3–6 months of additional delay per rejection cycle, depending on review turnaround times.

By omitting these delays and simplifying resubmission behavior, our model produces artificially shorter time-to-acceptance estimates and somewhat higher acceptance counts. The true disruption caused by AI-driven submission surges is therefore likely even greater than our reported results suggest.

### 5.4 AI Reviews and AI Triage

We also implicitly assumed in our model that journals would maintain their current acceptance capacity without increasing publication slots. This assumption is central to why our simulation produces such severe outcomes. If journals were able to expand their acceptance capacity in proportion to rising submissions, much of the backlog and extreme desk rejection rates we observed would be mitigated. Thus, if the "bad scenario" outlined in our models is to be avoided, this assumption would need to be relaxed.

---

[17] https://sakana.ai/ai-scientist-first-publication

However, increasing acceptance slots faces significant structural barriers. The most immediate constraint is the scarcity of qualified reviewers (Horta and Jung 2024; Thompson et al., 2024). Editors across disciplines already report that more than half of reviewer invitations are declined, and widespread "review fatigue" suggests that substantially expanding review volumes is unrealistic under current practices (Beecher and Wang, 2025). Even if additional papers could be reviewed, dramatically enlarging the set of accepted articles risks diluting journal reputation, since exclusivity and selectivity are core to journal prestige in business schools and other fields[18].

A more plausible adaptation is the adoption of AI-assisted triage and, eventually, AI-assisted reviewing (Checco et al. 2021). Automated tools could help editors quickly screen submissions for fit, completeness, or methodological red flags, reducing manual workload in the desk rejection stage. Similarly, AI could provide preliminary reviews or quality checks that lighten the burden on human reviewers (Wei et al., 2025), enabling some capacity expansion without eroding standards. While such systems raise their own ethical and quality-control questions, they may represent a viable path forward if journals are to absorb the exponential influx of AI-driven submissions without collapsing their review pipelines.

**5.5 Alternative Outlet for Scholarly Outputs**

In our assumption, all submissions are absorbed by journals in the market. This simplifying assumption also led to the bottleneck we observed in the simulation. By forcing all AI-boosted submissions into a fixed set of journals, we ignore alternative outlets such as preprint servers, conference proceedings, or emerging digital platforms, which could potentially absorb at least part of the surge (Jiang, 2025b). If the academic system is to cope with exponential growth in "submission-ready" manuscripts, finding new venues to absorb and circulate knowledge will be more important than suppressing submissions entirely.

One possible path is the expansion of preprint culture, which already plays a central role in fields like physics and computer science (Nature, 2024). Another is the rise of social media platforms and decentralized science (DeSci) chains as channels for scholarly dissemination (Ding et al., 2022; Jiang, 2025a). These outlets can provide rapid, large-scale exposure while shifting some of the evaluative burden to open peer commentary, community curation, or blockchain-based reputation systems. The challenge is not only how journals maintain their selectivity, but also how the broader ecosystem diversifies its channels of knowledge dissemination. If adapted thoughtfully, such approaches could relieve pressure on journals while making scholarship more inclusive, transparent, and adaptive in the AI era.

---

[18] https://www.forbes.com/sites/madhukarpai/2020/11/30/how-prestige-journals-remain-elite-exclusive-and-exclusionary/

## 5.6 Institutional Reforms on Faculty Evaluation System

In order to address the need for diversifying dissemination channels, institutional criteria for personnel evaluation must also evolve. If tenure and promotion continue to rely almost exclusively on counts of publications in a narrow set of prestigious journals, then the bottlenecks highlighted in our simulations will only intensify. Faculty will be forced to compete over a fixed and shrinking pool of acceptance slots, while alternative dissemination channels such as preprints, social media, or decentralized platforms remain undervalued.

This brings us directly to the assumption about tenure evaluation systems. Our model assumes that evaluation criteria remain static, privileging publication counts in established journals. Under this assumption, the influx of AI-driven submissions translates into a direct and measurable disadvantage for most faculty, especially junior colleagues. But if institutions expand their evaluative frameworks to include alternative markers of scholarly impact, such as openly disseminated preprints, community-reviewed outputs, or contributions documented on DeSci chains, then the consequences of AI disruption may be less severe.

## 5.7 From Early Adoption to Formal Recognition

Lastly, our simulation models assume that, apart from early adopters, most faculty continue to produce at historical levels of one or two papers per year. This assumption highlights the widening gap between adopters and non-adopters, but in practice the disadvantage could be partially mitigated as more faculty begin using AI tools across the research pipeline: from idea generation and literature review to data analysis and scientific writing. It is reasonable to expect that integration of AI into daily scholarly routines will accelerate in the near future (Taherdoost and Madanchian, 2024).

At present, the lack of formal recognition for AI-assisted scholarship makes adoption highly controversial. Faculty often confront dilemmas such as whether AI-assisted papers should be submitted at all, or whether the "trace" of AI in writing style should be deliberately minimized to avoid detection or stigma (Kwon, 2025). These uncertainties create not only anxiety but also incentives for non-transparency, where scholars conceal or downplay their use of AI rather than openly acknowledging it (van Niekerk et al., 2025). Until evaluation systems explicitly define what constitutes legitimate use of AI in research, adoption will remain uneven, and publication outcomes will continue to be shaped as much by cultural and ethical hesitations as by technical capabilities. In fact, I relied heavily on AI in writing this paper, including generating the code used for simulation. A significant part of AI-assisted research lies in critically verifying and refining AI outputs—a process I am still engaged in as this paper remains in preprint form.

Each of these assumptions shapes how we should interpret the outcomes. Thus, while the numbers in our tables are not literal predictions, they should be read as precautionary scenarios. They underscore the structural vulnerabilities of a tenure system tied to publication counts in a world where AI submissions can scale far beyond the carrying capacity of traditional journals.

# 6. Conclusion

This paper used simulation models to examine the potential impact of AI-driven manuscript surges on the academic publishing ecosystem, with particular attention to how these shifts may alter faculty research portfolios and tenure outcomes. Through a series of simulation models, including a Baseline model, an Early Adopter model, and an AI Abuse model, we explored how increases in submission volumes affect desk rejection rates, desk rejection times, review cycles, and eventual acceptance outcomes. The results consistently demonstrate that while early adopters of AI may temporarily gain an advantage by dramatically expanding the number of manuscripts ready for submission, the broader effect on the system is a sharp decline in acceptance opportunities for the majority of faculty. At high levels of AI adoption, even highly productive scholars face significant barriers, as desk rejection rates approach saturation and reviewer capacity becomes increasingly constrained.

## 6.1 Contributions to Academia

The study makes several contributions to the academic literature. First, it introduces simulation modeling as a methodological approach to studying academic publishing, a domain where empirical data on triage rates, review timelines, and editorial behavior are often opaque. While not intended as precise forecasts, the models provide a structured framework for analyzing systemic vulnerabilities and testing alternative scenarios. Second, the paper highlights the structural fragility of the publication system, particularly the desk rejection process, which emerges as the most immediate bottleneck under AI-driven submission shocks. Third, it situates the discussion of AI within the context of faculty evaluation and tenure systems, showing how reliance on publication counts may unfairly disadvantage junior colleagues in an environment where acceptance probabilities decline dramatically. By combining insights from both editorial practice and faculty career trajectories, the paper bridges the gap between publishing studies and higher education policy.

## 6.2 Implications for Practice

The findings carry several implications for the academic community. For faculty members, the results underscore the growing importance of strategic journal targeting, collaborative practices, and potentially transparent use of AI tools to enhance efficiency without compromising integrity. For journal editors and publishers, the simulations emphasize the need to explore AI-assisted triage and review support systems in order to prevent editorial bottlenecks from paralyzing the pipeline. Finally, for institutions and tenure committees, the results suggest that evaluation frameworks must adapt. Continued reliance on narrow publication counts risks penalizing otherwise productive faculty, particularly those entering the system during a period of heightened AI-driven competition. Alternative evaluative metrics, such as preprints, open peer commentary, or documented engagement in decentralized platforms or social media, could help rebalance fairness in an evolving landscape.

# 7. Limitations and Future Work

This study, however, is bounded by important limitations. The simulations rely on simplifying assumptions, such as fixed journal capacity, instantaneous resubmission, and static faculty productivity among non-adopters. These assumptions mean the results should be interpreted as upper-bound scenarios rather than literal predictions. Data availability also constrains precision: desk rejection rates and review times vary significantly across journals and fields, and our estimates combine official statistics, editor anecdotes, and crowdsourced reports. Moreover, ethical and cultural hesitations around AI use remain difficult to model but may substantially affect outcomes.

Future research should address these gaps in two ways. Empirically, surveys and collaborations with journal offices could provide more accurate estimates of triage rates, reviewer fatigue, and AI-related submissions, which will improve the usefulness of simulation models. Methodologically, future models can incorporate heterogeneous faculty strategies, such as cycling within tiers, delaying resubmissions, or strategically blending AI and human input. Finally, more attention should be given to alternative dissemination channels, including preprints, social media, and DeSci platforms, that may absorb excess submissions and reshape the criteria for academic recognition. By extending both the data and the models, scholars and institutions can better anticipate and adapt to the disruptive effects of AI on the publishing system.

Ding, W., Hou, J., Li, J., Guo, C., Qin, J., Kozma, R., & Wang, F. Y. (2022). DeSci based on Web3 and DAO: A comprehensive overview and reference model. *IEEE Transactions on Computational Social Systems*, *9*(5), 1563-1573.

Hanson, M. A., Barreiro, P. G., Crosetto, P., & Brockington, D. (2024). The strain on scientific publishing. *Quantitative Science Studies*, *5*(4), 823-843.

Horta, H., & Jung, J. (2024). The crisis of peer review: Part of the evolution of science. *Higher Education Quarterly*, *78*(4), e12511.

Huisman, J., & Smits, J. (2017). Duration and quality of the peer review process: the author's perspective. *Scientometrics*, *113*(1), 633-650.

Jansen, P., Tafjord, O., Radensky, M., Siangliulue, P., Hope, T., Mishra, B. D., ... & Clark, P. (2025). Codescientist: End-to-end semi-automated scientific discovery with code-based experimentation. *arXiv preprint arXiv:2503.22708*.

Jong, S., & Kantimm, W. (2024). Do professional staff in universities really challenge academic norms? A perspective from the Netherlands. *Higher Education*, *88*(6), 2165-2186.

Jiang, S. (2025a). MindStream Research Framework Concept Draft. Available at SSRN: https://ssrn.com/abstract=5389112 or http://dx.doi.org/10.2139/ssrn.5389112

Jiang, S. (2025b). Overcoming Echo Chambers: Harnessing Cross-Platform Feedback on Social Media. In Proceedings of the 7th New England Associatio of Information Systems (NEAIS2025).

Kim, J., Lee, Y., & Lee, S. (2025). Position: The AI Conference Peer Review Crisis Demands Author Feedback and Reviewer Rewards. *arXiv preprint arXiv:2505.04966*.

Kovanis, M., Trinquart, L., Ravaud, P., & Porcher, R. (2017). Evaluating alternative systems of peer review: a large-scale agent-based modelling approach to scientific publication. *Scientometrics*, *113*(1), 651-671.

Kwon, D. (2025). Is it OK for AI to write science papers? Nature survey shows researchers are split. *Nature*, *641*(8063), 574-578.

Launio, C. C., Samuel, F. K. D., Banez, A., Talastas, M. C., Sito, L., & Cruz, K. B. D. (2024). Motivating Factors and Challenges of Faculty Members in a State University in the Philippines in Publishing Journal Articles. *Journal of Scientometric Research*, *13*(3), 866-876.

Lee, C. J., Sugimoto, C. R., Zhang, G., & Cronin, B. (2013). Bias in peer review. *Journal of the American Society for information Science and Technology*, *64*(1), 2-17.

Lu, C., Lu, C., Lange, R. T., Foerster, J., Clune, J., & Ha, D. (2024). The ai scientist: Towards fully automated open-ended scientific discovery. *arXiv preprint arXiv:2408.06292*.

Miao, J., Davis, J. R., Pritchard, J. K., & Zou, J. (2025). Paper2Agent: Reimagining Research Papers As Interactive and Reliable AI Agents. *arXiv preprint arXiv:2509.06917*.

Moss, A. (2025). The AI Cosmologist I: An Agentic System for Automated Data Analysis. *arXiv preprint arXiv:2504.03424*.
24

# Appendix A. Normal Distributions for Review Cycles and Author Revision Time

**Proposed distributions**

**Review time per round (editor/reviewer side)**

We use a **truncated Normal** (never negative), centered at review cycle means:

- **T1:** $\mathcal{N}(\mu=6.0 \text{ mo}, \sigma=1.5 \text{ mo})$
- **T2:** $\mathcal{N}(\mu=5.0 \text{ mo}, \sigma=1.2 \text{ mo})$
- **T3:** $\mathcal{N}(\mu=4.0 \text{ mo}, \sigma=1.0 \text{ mo})$

**Author revision time per Major Revision**

We use a **truncated Normal** (never negative), centered at author revision means:

- **T1:** $\mathcal{N}(\mu=2.5 \text{ mo}, \sigma=0.8 \text{ mo})$
- **T2:** $\mathcal{N}(\mu=2.0 \text{ mo}, \sigma=0.6 \text{ mo})$
- **T3:** $\mathcal{N}(\mu=1.5 \text{ mo}, \sigma=0.5 \text{ mo})$

**Why these standard deviations?**

Based on SciRev data we collected, we estimated these values. It show wide dispersion in actual turnaround—reviewer availability, AE workload, seasonality, and paper complexity all swing timelines. A coefficient of variation ~20–30% for review cycles and ~25–35% for author revisions is reasonable and keeps extremes rare but possible. Higher-tier journals have more difficulty securing reviewers and typically request deeper changes, so we allow slightly larger σ for T1 than T2/T3.

# Appendix B Estimating Journal Acceptance Rates

A central parameter in the simulation is the probability of acceptance after review, which represents the likelihood that a manuscript sent to external referees will ultimately be accepted. Because most journals do not publish comprehensive statistics, and because acceptance rate reporting is inconsistent across fields, we rely on a combination of publicly available data from journal and publisher websites and field-specific logic grounded in business and management scholarship.

**Tier 1 (T1, A* / flagship journals)**

For Tier 1 (T1) journals—representing ABDC A* outlets and equivalent flagships—the acceptance probability is set at 0.09. This value is directly supported by acceptance rates reported on journal websites. These official figures provide credible anchors for a conservative baseline. Representative journals include:

- *Strategic Management Journal*: ~7% acceptance rate
- *Journal of Management Studies*: ~6% acceptance rate
- *Organization Science*: <10%, editorials note ~8%
- *Management Science*: ~11.8–12.1% depending on year
- *Academy of Management Journal* (AoM): ~8–10% (historical editorial reports)
  Source: AoM editorials.

Note that this is not a comprehensive list of A* journals and we only used journals which displayed officially available numbers on websites. Averaging across these journals yields values between 6–13%, with the median clustering around 8–10%. Hence T1 parameter was selected at 0.09.

**Tier 2 (T2, A / strong field journals)**

For Tier 2 (T2) journals—representing ABDC A outlets with strong disciplinary impact. Again, we draw from reported values on journals' websites. The average of these values, combined with the position of these journals as strong but not flagship outlets, justifies the 0.12 parameter choice. Representative journals include:

- *International Journal of Management Reviews* (Wiley): ~5%
  Source: Wiley Journal Insights.
- *Gender, Work & Organization* (Wiley): ~13%
  Source: Wiley Journal Insights.
- *European Journal of Operational Research* (Elsevier): ~13%
  Source: Elsevier journal metrics.

Strong A-level journals vary more widely than A*. The range is ~5–13%, with several clustered around 12–13%. The parameter was set at **0.12** to represent this midpoint.

**Tier 3 (T3, other journals)**

For Tier 3 (T3) journals, we do not list them here. The acceptance probability is set at 0.24. This estimate reflects two considerations. First, T3 covers a much broader set of journals than T1 and T2 combined, including ABDC B and C journals, specialty outlets, and developmental publications. Many of these journals explicitly aim to provide accessible platforms for authors who are less competitive at A*/A levels. Second, author submission behavior plays a role: when scholars seek quicker publication or are unable to place a paper in higher-tier outlets, they often target journals perceived as "easier" to publish in. This subjective selection bias inflates observed acceptance rates at the lower tiers, since authors self-select outlets where the probability of success is higher.

It is important to emphasize that these values represent overall acceptance probabilities, not strictly conditional acceptance given review. Because desk rejections are embedded in the reported figures, the estimates are conservative: the actual probability of acceptance after a full review round may be higher.

# Appendix C Review Model

**Model Assumptions**

- We only consider manuscripts sent out to reviews and do not consider desk rejection in this model.
- Let K = number of revision rounds required for acceptance.
- In round k = 1...K, the editor gives acceptance A with probability $a_k$, major revision MR with probability $m_k$, otherwise Reject with probability $(1-a_k-m)$. We don't model Minor Revision, considering it is very close to acceptance, and the time needed for minor revision is negligible.
- "Kind World "assumption: If the manuscript receives MR in *all* K rounds, it is automatically accepted (to avoid complicating the model too much).
- Acceptance rate $a_k$ increases across rounds.
- Major revision rate $m_k$ decreases across rounds.
- Then the overall acceptance C is (conditional on entering reviews, K=3 for example)

$$C = a_1 + m_1\Big(a_2 + m_2\big(a_3 + m_3\big)\Big)$$

- The eventual acceptance rate is C*(1-desk rejection rate)

**Baseline choice**

We use K = 3 major-revision rounds (common in practice). First round acceptance rates are set to low (T1: $a_1$=0.00; T2: $a_1$=0.01; T3: $a_1$=0.02). We target eventual acceptance rate for T1=0.09, T2=0.12, T3=0.24 (assumptions in previous Appendix). The the final calibrated A and MR probabilities are:

| Tier | Target: Eventual Accep. Rate | First Round Prob. (a, m, 1-a-m) | Second Round Prob. (a, m, 1-a-m) | Third Round Prob. (a, m, 1-a-m) | Overall accep. Rate C |
|---|---|---|---|---|---|
| T1 (A*) | 0.09 | 0.00, 0.55, 0.45 | 0.05, 0.50, 0.45 | 0.50, 0.45, 0.05 | ≈ 0.285 |
| T2 (A) | 0.12 | 0.01, 0.55, 0.44 | 0.10, 0.44, 0.46 | 0.40, 0.35, 0.25 | ≈ 0.247 |
| T3 (B/C) | 0.24 | 0.02, 0.55, 0.43 | 0.20, 0.37, 0.43 | 1.00, 0.00, 0.00 | ≈ 0.342 |

**Interpretation example**: a manuscript is sent to an A* journal. The manuscript survives desk rejection with $p_0$=40%. Conditional on being sent to reviews, the manuscript then has $a_1$=1% in immediate acceptance, $m_1$ =48% chance in receiving a major revision, and 51% in rejection in the first round of reviews.

It is important to note that the round-level probabilities of acceptance, revision, and rejection are not intended to precisely represent actual editorial practices. In reality, the exact a, v, and $m$ values vary across journals, evolve over time, and depend heavily on manuscript quality. Instead, these parameters serve as simulation inputs that capture the dynamics of editorial decision cycles in an aggregate sense. What matters for our purposes is how these probabilities shape broader outcomes: at the journal level, they determine rejection patterns—whether manuscripts are desk-rejected quickly or filtered out after extended review cycles. At the faculty level, they govern the distribution of publication counts, since longer or harsher review paths reduce the number of manuscripts that can successfully reach acceptance within a fixed career window. Thus, while the exact round-level parameters are heuristic, the eventual acceptance rate they produce is a robust and more interpretable measure, especially when evaluating how AI-driven submission surges may shift tenure-relevant publication profiles.